\documentclass[aps,prl,twocolumn,groupedaddress]{revtex4}

\usepackage{amssymb,amsmath}
\usepackage{graphicx}
\usepackage{ulem}
\usepackage{color}
\def\beq{\begin{equation}}
\def\eeq{\end{equation}}

\begin{document}

\title{Realization of a quantum walk with one and two trapped ions}

\author{F.~Z\"ahringer$^{1,2}$}
\author{G. Kirchmair$^{1,2}$}
\author{R.~Gerritsma$^{1,2}$}
\author{E.~Solano$^{3,4}$}
\author{R. Blatt$^{1,2}$}
\author{C. F. Roos$^{1,2}$}

\affiliation{$^1$Institut f\"ur Experimentalphysik, Universit\"at Innsbruck, Technikerstr.~25, A-6020 Innsbruck, Austria\\
$^2$Institut f\"ur Quantenoptik und Quanteninformation,
\"Osterreichische Akademie der Wissenschaften, Otto-Hittmair-Platz
1, A-6020 Innsbruck, Austria\\
$^3$ Departamento de Qu\'imica F\'isica, Universidad del Pa\'is Vasco - Euskal Herriko
Unibertsitatea, Apdo. 644, 48080 Bilbao, Spain\\
$^4$ IKERBASQUE, Basque Foundation for Science, Alameda Urquijo 36, 48011 Bilbao, Spain}

\date{\today}

\begin{abstract}
We experimentally demonstrate a quantum walk on a line in phase space using one and two trapped ions. A walk with up to 23 steps is realized by subjecting an ion to state-dependent displacement operations interleaved with quantum coin tossing operations. To analyze the ion's motional state after each step we apply a technique that directly maps the probability density distribution onto the ion's internal state. The measured probability distributions and the position's second moment clearly show the non-classical character of the quantum walk. To further highlight the difference between the classical (random) and the quantum walk, we demonstrate the reversibility of the latter. Finally, we extend the quantum walk by using two ions, giving the walker the additional possibility to stay instead of taking a step.
\end{abstract}

\pacs{03.67.Ac,37.10.Vz,42.50.Dv}
% 03.67.Ac 	Quantum algorithms, protocols, and simulations
% 37.10.Ty 	Ion trapping
% 37.10.Vz 	Mechanical effects of light on atoms, molecules, and ions
% 42.50.Dv 	Quantum state engineering and measurements

\maketitle

The Galton board~\cite{Galton:1889} is a mechanical device in which a falling ball encounters a triangular lattice of pins stuck in a board that repeatedly scatter the ball to the left or right in a random way. Originally conceived for illustrating the emergence of normal probability distributions, it can also be considered as an apparatus for carrying out a random walk on a line~\cite{Pearson:1905, Rayleigh:1905}, a notion that had not been introduced into the scientific literature at that time.

Since then, random walks have become an ubiquitous concept in physics and computer science. The quantum walk~\cite{Aharonov:1993,Kempe:2003} is the quantum analogue of a random walk. In its discrete one-dimensional version, a spin-$\frac{1}{2}$ quantum particle initially described by a wave packet centered at position $x_0$ undergoes a one-dimensional motion governed by the particle's internal state. The particle is state-dependently displaced by a step of length $d$ by the action of the unitary operator $U_d=\exp(-\frac{i}{\hbar}\sigma_j \hat{p}d)$ where $\sigma_j$ is a spin projection operator and $\hat{p}$ the momentum operator (see Fig.~1).
This operation is followed by another unitary operation $U_i=\exp(-i\frac{\pi}{4}\sigma_k)$  with $Tr(\sigma_j\sigma_k)=0$ scrambling the particle's internal state. After $N$ iterations of this elementary step, the particle's initial wave function $|\Psi_0\rangle$ has evolved into
\begin{equation}
|\Psi_N\rangle=\left(U_iU_d\right)^N\,|\Psi_0\rangle=\left(e^{-i\frac{\pi}{4}\sigma_k}e^{-\frac{i}{\hbar}\sigma_j \hat{p} d}\right)^N\,|\Psi_0\rangle.
\label{eq:QuantumWalk}
\end{equation}
After $N$ steps, for a wave packet initially localized at $x_0=0$, the wave packet is spread out over a distance $2Nd$. Moreover, due to quantum interference of different paths, the spatial probability distribution strongly differs from the classical case. While for the classical random walk a binomial probability distribution results with $\langle x^2 \rangle\propto N$, the distribution for the quantum walk is peaked towards the outer edge and has a second moment growing like $\langle \hat{x}^2 \rangle\propto N^2$.

\begin{figure}
\includegraphics[width=5cm]{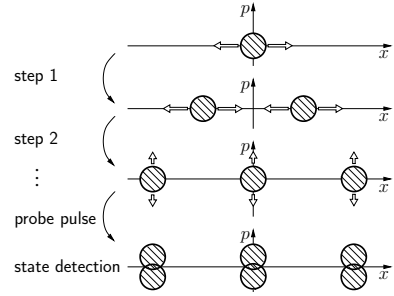}
\caption{\label{fig_Method} Quantum walk in phase space. In each step of the walk, a state-dependent displacement operation splits the wave function in phase space into two parts followed by a coin tossing operation that coherently scrambles the internal state of the ion. These operations are repeated $N$ times. To measure marginal distributions in phase space, a probe pulse is applied that state-dependently displaces the wave function in phase space in a direction orthogonal to the one to be measured.}
\end{figure}

There have been a number of proposals discussing experimental realizations of one-dimensional quantum walks in systems like atoms in optical lattices~\cite{Dur:2002a}, trapped ions~\cite{Travaglione:2002}, or cavity QED~\cite{Sanders:2003}. Recently, experimental realizations with atoms in an optical lattice~\cite{Karski:2009}, a trapped ion~\cite{Schmitz:2009a}, and photons~\cite{Schreiber:2009} have been reported.

For the case of trapped ions, different techniques for analyzing the quantum walk have been discussed~\cite{Travaglione:2002, Xue:2009} and a proof-of-principle experimental realization was reported recently~\cite{Schmitz:2009a} for a limited number of steps. In this paper, we demonstrate a discrete quantum walk with up to 23 steps using a single trapped ion and analyze it by a measurement technique that directly reconstructs the ion's probability density along a line in phase space.

In the experiment, a single $^{40}$Ca$^{+}$ ion is suspended in a linear Paul trap~\cite{Kirchmair:2009} with radial and axial trap frequencies of $\omega_{\rm r} \approx$ ($2\pi$) 3 MHz and $\omega_{\rm ax}$ = ($2\pi$) 1.356 MHz, respectively. Doppler cooling, resolved sideband cooling of the axial mode and optical pumping prepare the ion in the ground state of motion and the internal state $\lvert S_{ 1/2}, m = 1/2\rangle\equiv \lvert -\rangle_z$ \footnote{$\lvert \pm\rangle_k$ denotes the eigenstate of $\sigma_k$ with eigenvalue $\pm 1$.}. A narrow linewidth laser at 729~nm coherently couples the states $|-\rangle_z$ and $\lvert D_{5/2}, m = 3/2\rangle\equiv \lvert +\rangle_z$. State detection is done via fluorescence detection on the $S_{1/2}\leftrightarrow P_{1/2}$ transition~\cite{Kirchmair:2009}.

A general state-dependent displacement Hamiltonian is implemented using a bichromatic light field at 729~nm that is resonant with both the blue and red axial sideband of the $\lvert -\rangle_z\leftrightarrow\lvert +\rangle_z$ transition. In the Lamb-Dicke regime, the resulting Hamiltonian, which is the sum of a Jaynes-Cummings and an anti-Jaynes-Cummings Hamiltonian, is given by

\begin{align}\label{H_displace}
H_{D} =& \hbar\eta\Omega\left(\left(\sigma_{x}\cos{\phi_{+}} - \sigma_{y}\sin{\phi_{+}} \right)\right.\notag\\
&\ \otimes \left.\left((a + a^{\dagger})\cos{\phi_{-}} + i(a^{\dagger} - a)\sin{\phi_{-}}\right)\right).
%&\ \otimes \left(\hat{x}\cos{\phi_{\rm -}} + \hat{p}\sin{\phi_{\rm -}}\right))
\end{align}
Here, $\eta = 0.06$ is the Lamb-Dicke parameter, $\Omega$ the Rabi frequency and $2\phi_{+}=\phi_{\rm r}+\phi_{\rm b}$ and $2\phi_{ -}=\phi_{\rm b}-\phi_{\rm r}$ are the sum and the difference, of the phases of the light fields tuned to the red and blue sideband.

To perform a symmetric quantum walk the ion is prepared in the state $|+\rangle_y= (\lvert +\rangle_z + i\lvert -\rangle_z)/\sqrt{2}$ by a $\pi/2$-pulse on the carrier transition. Applying the bichromatic light field with $\phi_{-} = \pi/2$ and $\phi_{+} = 0$ realizes the Hamiltonian $H_{d} = 2 \eta \Omega \Delta_x \sigma_{x} \hat{p}$
with the momentum operator $\hat{p}=\frac{a^\dagger - a}{2}\frac{i\hbar}{\Delta_x}$ and $\Delta_x = \sqrt{\frac{\hbar}{2 m \omega_{\rm ax}}}$. Application of this Hamiltonian for a duration $\tau$ generates the propagator $U_d$ with step size $d=2\eta\Omega\tau\Delta_x$.

\begin{figure}[h!]
\includegraphics[width=8.5cm]{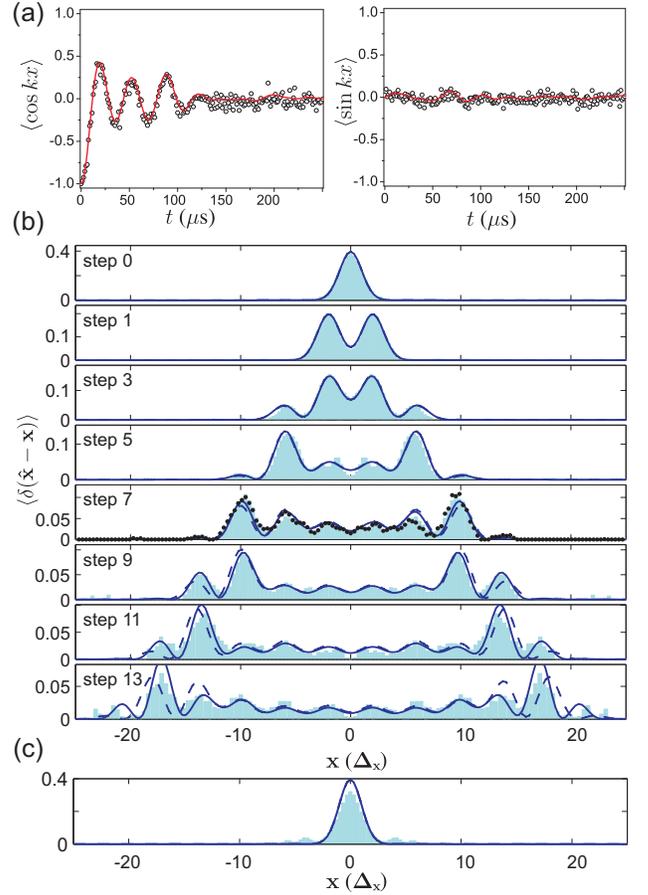}
\caption{\label{fig_Walk} (a) Measurement of Fourier components $\left<\cos(kx)\right>$ and $\left<\sin(kx)\right>$ for a seven-step quantum walk. The data are obtained by varying the duration of the probe pulse for the ion prepared in the internal state $|+\rangle_z$ (left) or $|+\rangle_y$ (right) after completing the walk. The probability distribution is obtained by Fourier transforming a fit to the data (solid line). (b) Reconstruction of the symmetric part of the probability distribution $\langle\delta(\hat{x}-x)\rangle$ for up to 13 steps in the quantum walk. The blue dashed curve is a numerical calculation for the expected distribution within the Lamb-Dicke regime. The blue solid curve takes into account corrections to the Lamb-Dicke regime. In step 7, the dotted curve represents the full reconstruction using also the $\left<\sin(kx)\right>$ shown in (a).(c) Probability distribution of a five-step quantum walk after application of five additional steps which invert the walk and bring it back to the ground state.}
\end{figure}

Under the action of $H_{d}$, the ion's wave packet coherently splits in phase space along the $x$-axis. The two emerging wave packets $\psi_1^{(m)}$, $\psi_2^{(m)}$ are associated with the internal states $|\pm\rangle_x$. The length and the intensity of the pulse determine the width of the splitting. In our experiments we use a pulse of 40~$\mu$s with a Rabi frequency of $\Omega= (2\pi)\,68$~kHz to achieve a step size of $d = 2\Delta_x$. This step size makes the two resulting motional wave packets nearly orthogonal, $|\langle\psi_1^{(m)}|\psi_2^{(m)}\rangle|^2\approx 0.02$, but still allows for a large number of steps in phase space. Next, we perform a $\pi/2$-pulse acting on the carrier transition as a symmetric coin flip. This pulse creates an equal superposition of $\sigma_x$ eigenstates for both wave packets. These two pulses are repeated according to the number of steps to be carried out.

\begin{figure}
\includegraphics[width=7cm]{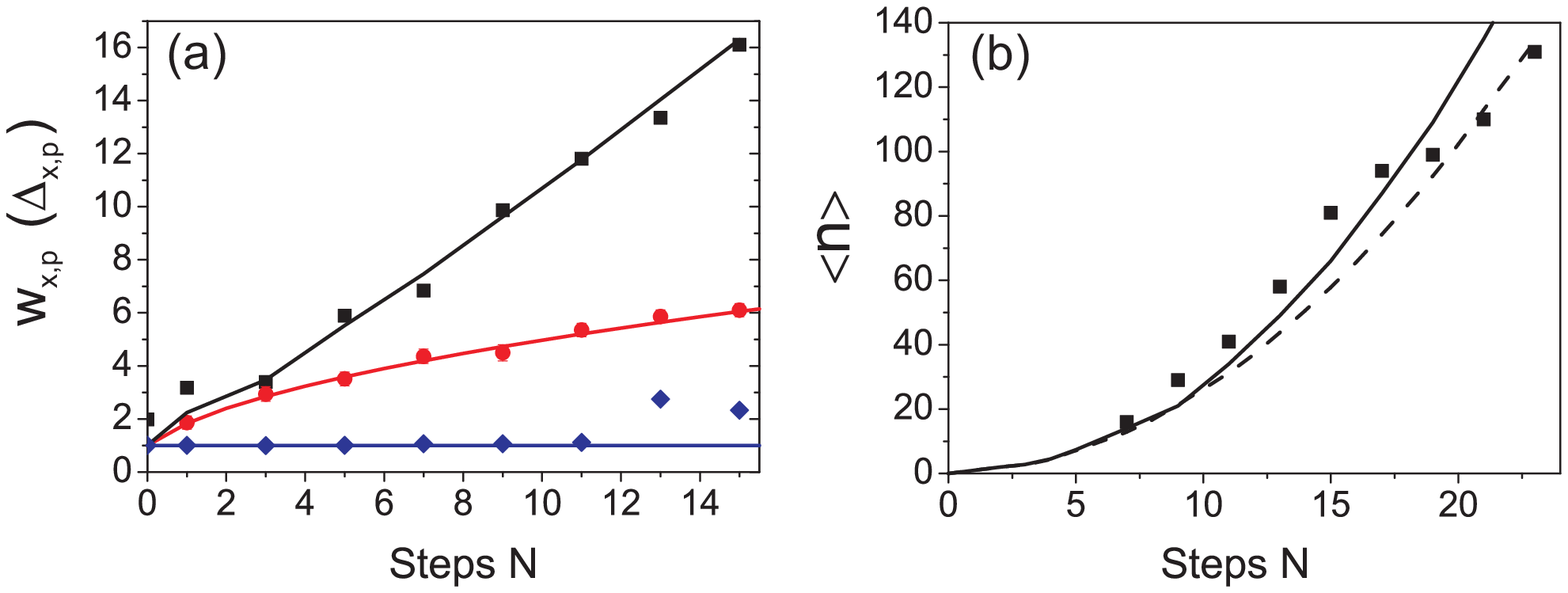}
\caption{\label{fig_Sigma}(Color online) (a) Width $w_x$ of the probability distribution in units of ground state size $\Delta_x$ as a function of the number of steps for a quantum (${\blacksquare}$) walk. The solid curve represents a full numerical simulation of the quantum walk as realized in the experiment. The width of the $x$-distribution for a classical random walk ($\textcolor[rgb]{1,0,0.00}{\bullet}$) increases more slowly and is described (solid red line) by eq.~(\ref{eqCl_sigma}). The data points (${\color{blue} \blacklozenge}$) show the measured width $w_p$ of the marginal distribution along the p-direction with $\Delta_p=\hbar/2\Delta_x$. (b) Average number of vibrational quanta after $N$ steps in the quantum walk measured by driving oscillations on the carrier transition. The solid line is based on a full simulation, the dashed line assumes the validity of the Lamb-Dicke approximation.}
\end{figure}

To measure the probability distribution along a line in phase space, we create two displaced copies of the state that are subsequently interfered. For this, we use of another state-dependent displacement operation $U_p=\exp(-ik\hat{x}\sigma_x/2)$~\cite{Wallentowitz:1995, Gerritsma:2010}.  A measurement of $\sigma_z$ following the application of $U_p$ is equivalent to measuring the observable
\beq\label{eqFourier}
O(k)=U_p^\dagger\sigma_z U_p=\cos(k\hat{x})\sigma_z+\sin(k\hat{x})\sigma_y,
\eeq
with the usual position operator $\hat{x}=(a^\dagger + a) \Delta_x$ on the initial state. The propagator $U_p$ is obtained by setting $\phi_+$ and $\phi_-$ in $H_{D}$ to 0. Here, $k=2\eta\Omega_p t/\Delta_x$ is proportional to the interaction time $t$. If the ion's internal state is $|+\rangle_z$, we have $\langle O(k)\rangle=\langle\cos(k\hat{x})\rangle$ and for $|+\rangle_y$, we have $\langle O(k)\rangle=\langle\sin(k\hat{x})\rangle$. A Fourier transformation of these measurements yields the probability density $\langle\delta(\hat{x}-x)\rangle$ in position space which for a pure state $|\Psi\rangle$ amounts to $|\Psi(x)|^2$. Furthermore, we have that $\left.\frac{d^2}{dk^2}\langle O(k)\rangle\right\vert_{t=0}\propto\langle\hat{x}^2\sigma_z\rangle$~\cite{Lougovski:2006}. For eigenstates of $\sigma_z$, the initial curvature of the expectation value $\langle O(k)\rangle$ thus gives the width of the probability distribution $w_x$.

The quantum walk entangles internal and motional degrees of freedom. Its analysis, however, requires the preparation of pure internal states like $|+\rangle_z$ or $|+\rangle_y$. Therefore, we recombine all internal state populations in $|-\rangle_z$ before the measurement. To this end, the population in $\lvert +\rangle_z$ is transferred to $\lvert -\rangle_z$ after transferring the population in $\lvert -\rangle_z$ to the auxiliary state $\lvert D_{\rm 5/2}, m = 5/2\rangle$. A laser pulse at 854~nm excites the population from $\lvert D_{\rm 5/2}, m = 5/2\rangle$ to $\lvert P_{\rm 3/2}, m = 3/2 \rangle$ from where it spontaneously decays to $\lvert -\rangle_z$. The efficiency of this pumping process is $>99$~\%, limited by a small branching ratio to the $D_{3/2}$-state. Only after the recombination step, we prepare the internal state required for measuring the even or odd Fourier components of (\ref{eqFourier}). Due to the small Lamb-Dicke parameter, the probability of changing the motional state of the ion during the pumping steps is small
and hardly affects measurements of observables in position space at all.
%, and could be eliminated completely by analyzing the motional state associated with the internal states $|+\rangle_z$ and $|-\rangle_z$ separately.

In the experiment we set $\Omega_{p}=(2\pi)\,26$~kHz and measure $\langle \sigma_z \rangle$ for probe times between 0 and 300~$\mu$s in order to reconstruct the probability distribution $\langle\delta(\hat{x}-x)\rangle$ for different numbers of steps $N$. Since the walk is symmetric, it is in principle sufficient to measure only the even components of~(\ref{eqFourier}). For a seven-step walk, the measured odd and even Fourier components are displayed in Fig.~\ref{fig_Walk}(a). Panel (b) shows the reconstructed probability distribution $\langle\delta(\hat{x}-x)\rangle$ based on the even components for up to 13 steps. The uneven terms were checked to be close to zero for each number of steps $N$. The dashed lines in the plots are numerical simulations based on the Lamb-Dicke approximation. These lines deviate from the reconstructed distribution for $N>7$ due to higher order terms in $\eta$ that are not taken into account in eq.~(\ref{H_displace}). The solid lines are based on a numerical simulation using all orders. A similar difficulty occurs in the measurement of observables based on eq.~(\ref{eqFourier}). For this reason, the reconstruction is not accomplished by a direct Fourier transformation of the data. Instead, we apply a constrained least-square fit based on convex optimization~\cite{cvx:2009} capable of handling higher-order corrections (see the EPAPS document for more information on the reconstruction process). To get smoother distributions additional constraints were invoked by the reconstruction algorithm. A physical constraint is given by the maximal kinetic energy a one-dimensional wave packet can have. An estimate for the kinetic energy can be determined by measuring the momentum distribution in the same way as the position distribution. By changing $\phi_{\rm -} $ to $\pi/2$ in the probe pulse, the operator $\hat{x}$ appearing in (\ref{eqFourier}) is replaced by an operator $\propto\hat{p}$. These measurements (see Fig.~\ref{fig_Sigma}(a)) indicate that the momentum distribution is not seriously affected during the walk, as expected for a pure displacement along the $x$-axis.

A striking difference between classical and quantum walks is the reversibility of the latter. In the experiment we reversed a quantum walk after five steps. This was done by switching the phase of the following five displacement and coin flipping pulses by $\pi$. In this way the quantum walk is exactly reversed and the ion returns to the ground state. The corresponding reconstructed probability distribution shown in Fig.~\ref{fig_Walk}(c) closely resembles the one of the initial state and demonstrates once more the coherence of the quantum walk.

To further highlight the differences between quantum and classical walks we also realized a classical walk by randomizing the phase between each step (while keeping the coin flip-displacement operator pair coherent for each individual step). The phase for each step was generated by a random noise generator. This mimics a completely mixed ensemble of measurement outcomes that behaves classically. A good way of quantifying the difference between the quantum and classical walks is by measuring the average width of the probability distributions. For a classical walk with a step size $d=s\Delta_x$ we have
\beq\label{eqCl_sigma}
w_x=\Delta_x\sqrt{\frac{2s^2N}{\pi}+1},
\eeq
where the second term takes into account the initial width $\Delta_x$ of the probability distribution. By contrast, for a quantum walk the width goes as $w_x \sim N$ for high $N$. To measure $w_x$ for the random walk, the curvature of $\langle \sigma_z \rangle$ at short probe-time was analyzed. Quadratic fitting gives direct access to the width $w_x$. For the quantum walk, $w_x$ was obtained from the measured probability distributions. In Fig.~\ref{fig_Sigma}(a) the results of these procedures can be seen seen for both a quantum and a classical walk.

\begin{figure}
\includegraphics[width=8.5cm]{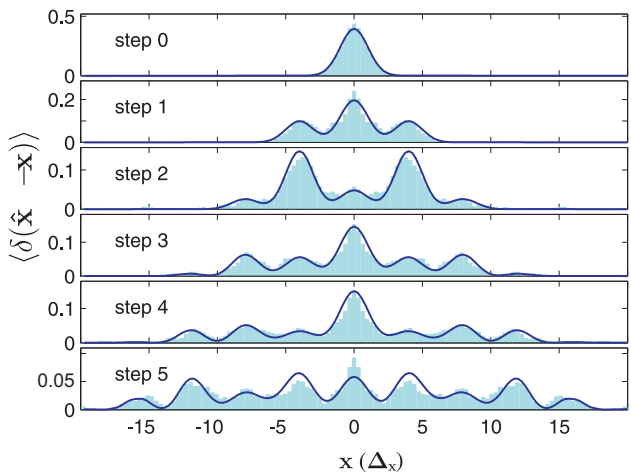}
\caption{\label{TwoIonWalk} Reconstructed probability distribution $\langle\delta(\hat{x}-x)\rangle$ for a two-ion quantum walk with up to 5 steps with a step size of $4\Delta_x$.}
\end{figure}

To avoid problems in the measurement of the motional state due to leaving the Lamb-Dicke limit for large numbers of steps, we implemented a method suggested in~\cite{Xue:2009}. Outside the Lamb-Dicke regime the coupling strength $\Omega_{n,n}$ on the carrier depends on the phonon number $n$ as $\Omega_{n,n} = \Omega_0 L_n(\eta^2)$. Here, $L_n(\eta^2)$ is the $n$-th order Laguerre polynomial. The mean phonon number $\left<n\right>$ is determined by a constrained least-square fit of the carrier Rabi flops with the number state distribution as a fit parameter. In Fig.~\ref{fig_Sigma}(b), the resulting average vibrational quantum numbers are shown. As expected for the quantum walk, we observe a quadratic dependence $\langle n\rangle \propto N^2$ on the number of steps.

Finally, we extend the quantum walk concept by adding a second ion to the system~\cite{Andraca:2005}. In the two ion quantum walk we make use of the center-of-mass mode. To account for the second ion, all Pauli matrices $\sigma_i$ in eq.~(\ref{eq:QuantumWalk}) are replaced by $\sigma_i^{(1)}+\sigma_i^{(2)}$. This changes the coin from two sided to four sided, with three possible operations. The "side" belonging to the state $|++\rangle_x$ ($|--\rangle_x$) corresponds to a step to the right (left) while the sides belonging to the states $|+-\rangle_x$ and $|-+\rangle_x$ correspond to no step at all. The ions are prepared in the state $|++\rangle_y$ with a $\pi/2$-pulse leading to a symmetric walk. For the two ion quantum walk all pulses are applied to both ions simultaneously. The probability distribution of the center of mass mode is obtained in the same way as for a single ion. The results for a walk of up to 5 steps are shown in Fig.~\ref{TwoIonWalk}. Again, the distribution deviates strongly from the classical version and shows a faster spreading.

In summary, we have implemented a quantum walk using trapped ions. An experimental technique was implemented to determine the probability distribution along a line in phase space. This method might have further applications in quantum optics experiments or quantum simulations~\cite{Gerritsma:2010}. We have highlighted the difference between a classical and a quantum walk and demonstrated the reversibility of the latter. The current limitation in number of steps is given by instabilities in the trap frequency leading to decoherence and by the change in the coupling strength due to high phonon numbers. Quantum walks are of importance as a primitive for quantum computation~\cite{Childs:2009} and in finding search quantum algorithms that outperform their classical counterparts~\cite{Hillery:2009}. As such the experimental implementation of the quantum walk serves as an important benchmark and points the way to further experiments. For instance, the implementation of a quantum walk with two ions opens up the interesting possibility to introduce entanglement~\cite{Andraca:2005} and more advanced walks.

\section*{Appendix: Reconstruction of the probability density}
For the reconstruction of the probability density $p(x)=\langle\delta(\hat{x}-x)\rangle$ of the motional quantum state $\rho_m$, we determine the expectation value of the observable
\beq\label{eqFourier}
O(k)=U_p^\dagger\sigma_z U_p=\cos(k\hat{x})\sigma_z+\sin(k\hat{x})\sigma_y
\eeq
by applying the unitary $U_p=\exp(-ik\hat{x}\sigma_x/2)$ to the state $\rho=|\Psi\rangle\langle\Psi|\otimes\rho_m$ and measuring the operator $\sigma_z$. The right-hand-side of eq.~(\ref{eqFourier}) is obtained by using the equality $\exp(i\theta\sigma_x)=\cos\theta\, I+i\sin\theta\,\sigma_x$ and $\sigma_i\sigma_j=\epsilon_{ijk}\sigma_k$ for $i\neq j$. In this way, we determine $\langle\cos(k\hat{x})\rangle$ by choosing $|\Psi\rangle=|+\rangle_z$, and $\langle\sin(k\hat{x})\rangle$ by $|\Psi\rangle=|+\rangle_y$ \cite{Wallentowitz:1995}. In principle, a Fourier transformation of $f(k)=\langle\cos(k\hat{x})\rangle+i\langle\sin(k\hat{x})\rangle$ is sufficient for obtaining the density $\langle\delta(\hat{x}-x)\rangle$ in position space. However, with a finite number of experiments, the expectation values can only be determined for a discrete number of $k$-values and these measurements do not yield the exact expectation values but rather estimates of them. As a consequence, a reconstruction of the density based on Fourier transformation gives unphysical probability densities that are not non-negative everywhere.
To overcome this problem, we reconstruct $p(x)$ by a constrained least-square optimization based on convex optimization~\cite{cvx:2009}. We discretize the position space by using a suitable set of points $x_i$ and search among the probability distributions with $p(x_i)\ge 0$ for all $x_i$ the distribution that minimizes
\begin{eqnarray}
\label{eqLeastSquare}
S&=&\sum_k\left(\sum_ip(x_i)\cos(kx_i)-C_k\right)^2\nonumber\\
&&+\sum_k\left(\sum_ip(x_i)\sin(kx_i)-S_k\right)^2
\end{eqnarray}
where $C_k$ and $S_k$ are the experimentally determined estimates of $\langle\cos(k\hat{x})\rangle$ and$\langle\sin(k\hat{x})\rangle$, respectively.

In our reconstruction, we use another physical constraint based on a measurement of the kinetic energy $\langle\frac{\hat{p}^2}{2m}\rangle$
%$\langle{p}^2/(2m)\rangle$.
in the following way. For a wave function $\psi=A(x)e^{i\phi(x)}$, a lower bound on the kinetic energy is given by
\begin{eqnarray}
\langle\frac{\hat{p}^2}{2m}\rangle&=&\frac{\hbar^2}{2m}\int_{-\infty}^\infty dx ((A(x)\phi^\prime(x))^2+A^\prime(x)^2)\nonumber\\
&\ge& \frac{\hbar^2}{2m}\int_{-\infty}^\infty dx A^\prime(x)^2
\end{eqnarray}
where differentiation with respect to $x$ is indicated by primes. For $p(x)=|\psi(x)|^2$, we then have because of $A(x)=p(x)^{\frac{1}{2}}$ and $A^\prime(x)=\frac{1}{2}p^\prime(x)p(x)^{-\frac{1}{2}}$ that
\beq
\label{eq:KineticEnergyConstraint}
\langle\frac{\hat{p}^2}{2m}\rangle\ge \frac{\hbar^2}{8m}\int_{-\infty}^\infty dx \frac{p^\prime(x)^2}{p(x)}.
\eeq
This constraint is also valid for mixed quantum states. In our optimization algorithm, eq.~(\ref{eq:KineticEnergyConstraint}) is a convex constraint that excludes distributions $p(x)$ having excessive energies. It requires a measurement of $\langle \hat{p}^2\rangle$ which is obtained by setting $\phi_-=\pi/2$ in the bichromatic Hamiltonian generating the unitary $U_p$ in eq.(\ref{eqFourier}) and calculating $d^2/dk^2\langle O(k) \rangle$. Adding the constraint works particularly well for the states produced by the quantum walk. These states ideally do not have any phase gradients $\phi^\prime(x)$ in which case inequality (\ref{eq:KineticEnergyConstraint}) turns into an equality.

Moreover, the optimization algorithm can also handle to some extent problems related to the validity of the Lamb-Dicke approximation. In this approximation, the bichromatic laser-ion interaction, which is used for the operation $U_p$ in the reconstruction measurement, is described by the Hamiltonian
\beq
\label{H_displace}
H_D = \hbar\eta\Omega\,\sigma_{x}\otimes (a + a^{\dagger}).
\eeq
This Hamiltonian is strictly valid only for $\eta\rightarrow 0$ because it is based on a Taylor expansion $e^{i\eta(a+a^\dagger)}=I+i\eta(a+a^\dagger)+{\cal O}(\eta^2)$ of the atom-light interactions that neglects terms in $\eta$ of order two or higher. If resonant terms up to third order are taken into account, the Hamiltonian becomes
\beq
\label{H_displace_thirdorder}
H_D = \hbar\eta\Omega\,\sigma_{x}\otimes \left[(a + a^{\dagger})-\frac{\eta^2}{4}\left((a+a^\dagger)\hat{n}+\hat{n}(a+a^\dagger)+1\right)\right].
\eeq
Since this Hamiltonian no longer commutes with $\hat{x}$, it cannot be used instead of eq.~(\ref{H_displace_thirdorder}) for the reconstruction procedure described above. On the other hand, an analysis based on eq.~(\ref{H_displace}) yields wrong results for quantum walks for large number of steps where the created states no longer fulfil the Lamb-Dicke criterion. Fortunately, for these states, their potential energy $\propto\langle(a+a^\dagger)^2\rangle$ is much larger than their kinetic energy $\propto\langle(i(a-a^\dagger))^2\rangle$ which provides the justification for replacing $\hat{n}$ by $\frac{1}{4}(a+a^\dagger)^2$ in eq.~(\ref{H_displace_thirdorder}). Using this assumption, the Hamiltonian
\beq
\label{H_displace_thirdorder_v2}
H_D = \hbar\eta\Omega\,\sigma_{x}\otimes \left[(a + a^{\dagger})(1-\frac{\eta^2}{8}((a+a^\dagger)^2+1))\right]
\eeq
becomes again diagonal in the real space basis. The reconstructed probability densities $\langle\delta(\hat{x}-x)\rangle$ shown in the paper are based on this Hamiltonian.

\begin{acknowledgments}
We gratefully acknowledge support by the Austrian Science Fund (FWF), by the
European Commission (Marie-Curie program), by the Institut
f\"ur Quanteninformation GmbH. E.S. acknowledges support from UPV-EHU GIU07/40, Ministerio de Ciencia e Innovaci\'on FIS2009-12773-C02-01, EuroSQIP and SOLID European projects.
This material is based upon work supported in part by
IARPA.
\end{acknowledgments}


\begin{thebibliography}{0}

\bibitem{Galton:1889} F.~Galton, {\it Natural inheritence} (Macmillan, 1889).

\bibitem{Pearson:1905} K. Pearson, Nature {\bf 72}, 294 (1905).

\bibitem{Rayleigh:1905} Rayleigh, Nature {\bf 72}, 318 (1905).

\bibitem{Aharonov:1993} Y. Aharonov, L. Davidovich, and N. Zagury, Phys.~Rev.~A {\bf 48}, 1687 (1993).

\bibitem{Kempe:2003} J. Kempe, Cont.~Phys. {\bf 44}, 307 (2003).

\bibitem{Dur:2002a} W. D\"ur, R. Raussendorf, V. M. Kendon, and H.-J. Briegel, Phys.~Rev.~A {\bf 66}, 052319 (2002).

\bibitem{Travaglione:2002} B. C. Travaglione and G. J. Milburn, Phys.~Rev.~A {\bf 65}, 032310 (2002).

\bibitem{Sanders:2003} B. C. Sanders, S. D. Bartlett, B. Tregenna, and P. L. Knight, Phys.~Rev.~A {\bf 67}, 042305 (2003).

\bibitem{Karski:2009} M. Karski {\it et al.},
% L. F\"orster, J.-M.Choi, A. Steffen, W. Alt, D. Meschede, and A. Widera,
Science {\bf 325}, 174 (2009).

\bibitem{Schmitz:2009a} H. Schmitz {\it et al.},
%R. Matjeschk, C. Schneider, J. Glueckert, M. Enderlein, T. Huber, and T. Schaetz,
Phys.~Rev.~Lett. {\bf 103}, 090504 (2009).

\bibitem{Schreiber:2009} A. Schreiber {\it et al.},
%K.-N. Cassemiro, V. Potocek, A. Gabris, P. Mosley, E. Andersson, I. Jex, and C. Silberhorn,
Phys. Rev. Lett. {\bf 104}, 050502 (2010).

\bibitem{Xue:2009} P. Xue, B. C. Sanders, and D. Leibfried, Phys.~Rev.~Lett. {\bf 103}, 183602 (2009).

\bibitem{Kirchmair:2009} G. Kirchmair {\it et al.},
%J. Benhelm, F. Z\"ahringer, R. Gerritsma, C. F. Roos, and R. Blatt,
New~J.~Phys. {\bf 11}, 023002 (2009).

\bibitem{Wallentowitz:1995} S. Wallentowitz and W. Vogel, Phys.~Rev.~Lett. {\bf 75}, 2932 (1995).

\bibitem{Gerritsma:2010} R. Gerritsma {\it et al.},
% G. Kirchmair, F. Z\"ahringer, E. Solano, R. Blatt, and C. F. Roos,
Nature {\bf 463}, 68 (2010).

\bibitem{Lougovski:2006} P. Lougovski, H. Walther, and E. Solano, Eur.~Phys.~J.~D {\bf 38}, 423 (2006).

\bibitem{cvx:2009} M. Grant and S. Boyd, CVX: Matlab software for disciplined convex programming (2009),
%\urlprefix
\url{http://stanford.edu/~boyd/cvx}.

\bibitem{Andraca:2005} S. E. Venegas-Andraca, J. L. Ball, K. Burnett, and S. Bose, New.~J.~Phys. {\bf 7}, 221 (2005).

\bibitem{Childs:2009} A. M. Childs, Phys.~Rev.~Lett. {\bf 102}, 180501 (2009).

\bibitem{Hillery:2009} M. Hillery, D. Reitzner, and V. Buzek, arXiv:0911.1102 (2009).
\end{thebibliography}
\end{document}